\documentclass[%
 reprint,
nofootinbib,
 amsmath,amssymb,
 aps,
]{revtex4-2}

\usepackage{graphicx}% Inclufigure files
\usepackage{dcolumn}% Align table columns on decimal point
\usepackage{bm}% bold math
\usepackage{hyperref}% add hypertext capabilities
\usepackage{color}
\usepackage[dvipsnames]{xcolor}
\usepackage{placeins}
\usepackage{multirow}

\hypersetup{colorlinks=true, citecolor=MidnightBlue,linkcolor=CornflowerBlue, urlcolor=CornflowerBlue, linktocpage=true}

\begin{document}

\preprint{APS/123-QED}

\title{
Robustness of kinetic screening against matter coupling
}% Force line breaks with \\
\author{Guillermo Lara}
\author{Miguel Bezares}
\author{Marco Crisostomi}
\author{Enrico Barausse}

\affiliation{SISSA, Via Bonomea 265, 34136 Trieste, Italy and INFN Sezione di Trieste}
\affiliation{IFPU - Institute for Fundamental Physics of the Universe, Via Beirut 2, 34014 Trieste, Italy}
\date{\today}% It is always \today, today,

\begin{abstract}

\noindent We investigate neutron star solutions in scalar-tensor theories of gravity with first-order derivative self-interactions in the action and in the matter coupling.
We assess the robustness of the kinetic screening mechanism present in these theories against general conformal couplings to matter. The latter
include ones leading to the classical Damour-Esposito-Far{\`e}se scalarization, 
as well as ones
depending on the kinetic term of the scalar field. 
We find that kinetic screening always prevails over scalarization, and that kinetic couplings with matter enhance the suppression of scalar gradients inside the star even more, without relying on the non-linear regime.
Fine tuning the kinetic coupling with the derivative self-interactions in the action allows one to partially cancel the latter, resulting in a weakening of kinetic screening inside the star. This effect represents a novel way to break screening mechanisms inside matter sources, and provides new signatures that might be testable with astrophysical observations.
\end{abstract}

\maketitle

%\tableofcontents

%%%%%%%%%%%%%%%%%%%%%%%%%%%%%%%%%%%%%%%%%%%%%%%%%%%%%%%%%%%%%%%%
%%%%%%%%%%%%%%%%%%%%%%%%%%%%%%%%%%%%%%%%%%%%%%%%%%%%%%%%%%%%%%%%
%%%%%%%%%%%%%%%%%%%%%%%%%%%%%%%%%%%%%%%%%%%%%%%%%%%%%%%%%%%%%%%%

\allowdisplaybreaks

\section{Introduction} \label{sec: Introduction}

Scalar-tensor theories propagate an additional scalar degree of freedom with respect to General Relativity (GR), which modifies the gravitational interaction at all scales.
Although such modifications may be welcome at cosmological scales, so as to account for the current accelerated expansion of the Universe \cite{Crisostomi:2017pjs}, on local scales they are strongly constrained by Solar System 
and binary pulsar tests of gravity~\cite{Damour:1991rd, Will:2014kxa, Will:1993hxu}.
One possibility to avoid such a clash with local tests of gravity, which are in strong agreement with GR, is that non-linear (self-)interactions suppress the scalar (fifth) force near matter sources, making gravity  locally similar to GR.
Starting from the most general, degenerate, higher-order scalar-tensor theories~\cite{Langlois:2015cwa, Crisostomi:2016czh,BenAchour:2016fzp}, consistency with the speed of gravitational waves (GWs) measured by GW170817~\cite{Monitor:2017mdv,TheLIGOScientific:2017qsa}, absence of GW decay into dark energy  \cite{Creminelli:2018xsv,Creminelli:2019nok}, and non-linear stability of the propagating scalar mode \cite{Creminelli:2019kjy} reduce the viable Lagrangian to the following form
\begin{eqnarray}
&& S=\!\!\int d^4 x \sqrt{-g} \Bigg[\Phi \,R +K(\varphi,X) \label{Jframe_action} \\
&& \qquad\qquad\qquad +\frac{6 \Phi_X^2}{\Phi} \varphi^{\mu} \varphi_{\mu \rho} \varphi^{\rho \nu} \varphi_{\nu}\Bigg] + S_m[g_{\mu\nu},\Psi_m]\,, \nonumber
\end{eqnarray}
where $R$ and $g$ are the Ricci scalar and metric determinant,
$\varphi$ is the scalar field (with $\varphi_\mu\equiv \partial_\mu\varphi, \, \varphi_{\mu\nu}\equiv \nabla_\nu\partial_\mu\varphi, \, X\equiv \varphi_\mu\varphi^\mu$), and $\Psi_m$ collectively describes the matter degrees of freedom.
In the above expression, $\Phi$ and $K$ are generic free functions of $\varphi$ and $X$ (with subscripts denoting partial derivatives, e.g. $\Phi_X \equiv \partial \Phi/\partial X$), and we have set $\hbar=c=1$.
Performing a conformal transformation $g_{\mu\nu} \to \Phi^{-1}\, g_{\mu\nu}$ 
from the Jordan frame
to the  Einstein frame, together with a redefinition of the free function $K$,
the action can be written as
 \begin{equation}
%\label{einframe}\!
\!S=\!\!\int\!d^4 x \sqrt{-g} \Bigg[ \frac{M_{\rm Pl}^2}{2}R + K(\varphi,X)
\Bigg] \! +S_m\!\!\left[\frac{g_{\mu\nu}}{\Phi(\varphi,X)},\Psi_m\right]~,
\label{eq: kessence action}
\end{equation}
where we have introduced the (reduced) Planck mass 
\(M_\mathrm{Pl}^{-2}=8\pi G\).
The corresponding theory is usually referred to as \(k\)-essence~\cite{Chiba:1999ka, Armendariz-Picon:2000nqq}.
For the rest of the paper 
we will work in the Einstein frame and will denote quantities in the Jordan frame with a tilde hat (e.g. $\tilde g$).

We assume that there are only two energy scales involved in the action, one associated with the metric and the scalar field ($M_\text{Pl}$) and the other with the derivatives of the scalar field ($\Lambda$). Moreover, we will assume a shift symmetry for the action (\(\varphi \to \varphi \,+ \text{const.}\)) which is only softly broken by a Planck suppressed scalar-matter interaction. Under these assumptions, we will only consider the lowest order non-linear terms in the free functions $K$ and $\Phi$, which read
\begin{align}
    K\left(X\right) &= -\dfrac{1}{2} X + \beta \dfrac{X^2}{4\Lambda^4} \,, \label{eq: Kes definition} \\
    \log \Phi \left(\varphi, X\right) &= \alpha_{1} \dfrac{\varphi}{M_\text{Pl}} + \alpha_{2} \dfrac{\varphi^2}{M_\text{Pl}^2}
    + \lambda_{1} \dfrac{X}{2\Lambda^4} 
    + \lambda_{2} \dfrac{X^2}{4\Lambda^{8}} \,, \label{eq: Phi definition}
\end{align}
where $\beta, \alpha_i, \lambda_i$ are \(\mathcal{O} \left(1\right)\) dimensionless constants and the $\log$ function
ensures $\Phi>0$ and thus the same metric signature in the Einstein and Jordan frames.
In particular, Fierz-Jordan-Brans-Dicke (FJBD) theory~\cite{Fierz:1956zz, Jordan:1959eg, Brans:1961sx} is recovered in the case\footnote{
Strictly speaking, FJBD theory corresponds to \(\beta = \lambda_{1, 2} = 0\).
However, in this paper we use this terminology also for cases where \(\lambda_{1, 2}\neq 0\).
}
\(\beta = 0\).
In the non-linear regime, one might be concerned about higher powers of $X$ becoming important, and whether it would 
be justified to neglect them. Also, higher derivative terms (with more than one derivative per field) may be generically expected to be produced by quantum corrections. Note however that Ref.~\cite{deRham:2014wfa} (without gravity) and later Ref.~\cite{Brax:2016jjt} (in presence of gravity) have shown that when radiative corrections are correctly computed, any given form of $K(X)$ is radiatively stable in the non-linear regime. Although there is no formal proof that this property holds  for the matter coupling functional form $\Phi(\varphi, X)$, the fact that we can eliminate this term through a conformal transformation to the Jordan frame is a good indication that this can be expected to be the case.

It is well known that when \(\beta < 0\) and 
\(\log \Phi\) 
depends only linearly on $\varphi$, a screening mechanism referred to as \(k\)-mouflage (or kinetic screening)~\cite{Babichev:2009ee}
suppresses the scalar force in the vicinity of a massive body, allowing for passing Solar System tests without imposing any bound on $\alpha_1$ \cite{Babichev:2009ee, Brax:2014gra, terHaar:2020xxb, Bezares:2021yek}. 
Furthermore, time evolutions 
for (cubic) \(k\)-essence 
have been shown to be well-posed \cite{Bezares:2020wkn, Bezares:2021yek} and numerical
simulations of merging binary neutron stars with screening have been performed in Ref.~\cite{Bezares:2021dma}.

On the contrary, in the absence of screening (i.e. $\beta \geq 0$), the same local tests constrain 
$|\alpha_1| < 5 \times 10^{-3}$
\cite{Will:2014kxa, Bertotti:2003rm}.
This bound has led to investigating the effect of quadratic corrections in $\varphi$ in the matter coupling, and to the discovery of spontaneous scalarization when
$\alpha_2 \gtrsim 2 $~\cite{Damour:1992we, Damour:1993hw}.

It is however unknown what  the effect of this term is in the presence of screening.
In principle, scalarization may  still occur
-- as the quadratic term
in $\varphi$ 
in the matter coupling still tends to render the scalar tachyonically unstable~\cite{Damour:1992we, Damour:1993hw,Andreou:2019ikc} -- and spoil the suppression of the scalar force. 
Likewise,  the effect of $X$-dependent couplings with matter [c.f. Eq.~(\ref{eq: Phi definition})] on screened solutions is currently unknown.
The aim of this work is to investigate these questions and assess the robustness of kinetic screening against matter couplings.

%%%%%%%%%%%%%%%%%%%%%%%%%%%%%%%%%%%%%%%%%%%%%%%%%%%%%%%%%%%%%%%%

This paper is organized as follows.
In Sec.~\ref{sec: theory}, 
we present the covariant equations of motion for the class of models described by Eqs.~\eqref{eq: Kes definition} and \eqref{eq: Phi definition}. 
In Sec.~\ref{sec: numerical integration}, we describe our methodology to obtain fully relativistic static solutions through numerical integration of the equations of motion in spherical symmetry.
In addition, in Sec.~\ref{sec: simplified analytic model}, we introduce an analytic model that captures the main features of the scalar configurations.
Our main results are presented in Sec.~\ref{sec: results}. 
Finally, in Sec.~\ref{sec: Conclusion}, we summarize our findings and draw our conclusions.
Details regarding the derivation of the equations of motion are relegated to Appendix~\ref{sec: app EOMs}.
Throughout this paper we use the \((-+++)\) signature for the metric.

%%%%%%%%%%%%%%%%%%%%%%%%%%%%%%%%%%%%%%%%%%%%%%%%%%%%%%%%%%%%%%%%
%%%%%%%%%%%%%%%%%%%%%%%%%%%%%%%%%%%%%%%%%%%%%%%%%%%%%%%%%%%%%%%%
%%%%%%%%%%%%%%%%%%%%%%%%%%%%%%%%%%%%%%%%%%%%%%%%%%%%%%%%%%%%%%%%

%
\section{Field equations } \label{sec: theory}

Variation of action \eqref{eq: kessence action} with respect to \(g_{\mu\nu}\) gives the equations of motion for the metric, 
\begin{align} \label{eq: Einstein equations}
   M_\text{Pl}^2 G_{\mu\nu} = T^{(\varphi)}_{\mu\nu}  + T_{\mu\nu} \,,
\end{align}
where \(G_{\mu\nu}\) is the Einstein tensor, 
the stress-energy tensor of the scalar field is given by
\begin{align} \label{eq: stress-energy tensor kessence definition}
    T^{(\varphi)}_{\mu\nu} = K(X) \, g_{\mu\nu} - 2 K_X (X) \varphi_\mu \varphi_\nu ~,    
\end{align}
and the matter stress-energy tensor is defined by
\begin{align}
    T^{\mu\nu} \equiv \dfrac{2}{\sqrt{-g}} \dfrac{\delta S_m}{\delta g_{\mu\nu}} ~,
\end{align}
with trace \(T \equiv g_{\mu\nu}T^{\mu\nu}\).

Variation with respect to 
\(\varphi\)
and the contracted Bianchi identity applied to Eq.~\eqref{eq: Einstein equations} give rise to the scalar and matter equations of motion 
(see Appendix \ref{sec: app EOMs} for more details), 
\begin{align} 
    \label{eq: scalar equation}
    \nabla_\mu \left(\mathcal{K} \, \varphi^\mu \right) &= \dfrac{1}{2} \mathcal{A} T ~, \\
    \label{eq: matter equation}
    \nabla_\mu \mathcal{T}^{\mu\nu} &= - \dfrac{\mathcal{T}}{2 \Phi} \nabla^{\nu} \Phi ~,
\end{align}
where we have defined
\begin{align} \label{eq: definition Q}
    \mathcal{K} &\equiv K_X + \mathcal{B} T ~, \\
    \label{eq: Tcurly definition}
    \mathcal{T}^{\mu\nu} &\equiv \Phi^{-3} \tilde{T}^{\mu\nu} = T^{\mu\nu} + 2 \mathcal{B} T \varphi^\mu \varphi^\nu ~,
\end{align}
with trace $\mathcal{T} \equiv g_{\mu\nu}\mathcal{T}^{\mu\nu}$, 
\begin{equation} \label{eq: A B C definitions}
    \mathcal{A} \equiv \dfrac{\Phi_\varphi}{\mathcal{C}}\,, \quad
    \mathcal{B} \equiv \dfrac{\Phi_X}{\mathcal{C}}\,, \quad
    \mathcal{C} \equiv -2 \Phi \left( 1 + \dfrac{X \, \Phi_X}{\Phi} \right)\,,
\end{equation}
and the Jordan-frame tensor is defined by
\begin{equation}
\tilde{T}^{\mu\nu} \equiv \frac{2}{\sqrt{-\tilde{g}}} \frac{\delta S_m}{\delta \tilde{g}_{\mu\nu}} \,.
\end{equation}

Notice in particular that 
the dependence of $\Phi$ on $X$ produces a matter redressing of $K_X$ in the scalar field equation and a disformal modification of the matter stress-energy tensor.

\section{Methodology} \label{sec: methodology}
\subsection{Numerical integration}\label{sec: numerical integration}

We describe matter as a perfect fluid in the Jordan frame by
\begin{align}
    \tilde{T}^{\mu\nu} =\left( \tilde{\rho} + \tilde{P}\right) \tilde{u}^{\mu}\tilde{u}^{\nu} + \tilde{P} \tilde{g}^{\mu\nu},
\end{align}
with trace \(\tilde{T} \equiv \tilde{g}_{\mu\nu}\tilde{T}^{\mu\nu}\),
where \(\tilde{u}^{\mu}\) is the 4-velocity of the fluid (normalized to \(-1\)) and the energy density \(\tilde{\rho} = \tilde{\rho}_0 \left(1 + \tilde{\epsilon}\right)\) is given in terms of the rest-mass density \(\tilde{\rho}_0\) and the internal energy \(\tilde{\epsilon}\).

We restrict to static solutions in spherical symmetry
and
write the line element in polar coordinates as
\begin{align}
    ds^2 = - N^2(r) \, dt^2 + a^2(r) \, dr^2 + r^2 d \Omega^2 ~,
\end{align}
where \(d \Omega^2 = d\theta^2 + \sin^2\theta d \phi^2\).
In addition to the scalar and matter equations, we use the 
\(tt\)- and \(rr\)-components
of the Einstein equations~\eqref{eq: Einstein equations}.
We will mainly focus on neutron star matter, and thus we close the system by specifying a polytropic equation of state (EOS)
\(\tilde{P} = \tilde{K} \tilde{\rho}_{0}^{\tilde{\Gamma}}\)
and \(\tilde{P} = (\tilde{\Gamma}-1)\tilde{\rho}_0 \tilde{\epsilon}\),
with adiabatic index \(\tilde{\Gamma} = 2\)
and 
\(\tilde{K} = 123\, G^3 M^2_\odot \)
 in the Jordan frame. 
 It is worth noticing that the EOS considered in this work is an approximate model for cold neutron stars~\cite{RezzollaZanotti2013rehy.book.....R}, therefore suitable for static solutions. At the same time, polytropic models display consistent mass-radius relations with observational constraints~\cite{Ozel:2016oaf}.

The final
Tolman–Oppenheimer–Volkoff (TOV) equations are a set of ordinary differential equations with schematic form
\begin{align} \label{eq: TOV schematic form}
    \boldsymbol{U}' = \boldsymbol{V}\left[\boldsymbol{U} , r\right]~,
\end{align}
where \(\boldsymbol{U} = (\varphi, \varphi', \tilde{P}, N, a)\) and the prime denotes the radial derivative. We do not write explicitly these equations as they are cumbersome and not particularly illuminating.

Regularity at the center of the star is imposed by solving the equations perturbatively around \(r = 0\). In this way, the resulting independent integration constants are found to be \(\{N(0), \varphi(0), \tilde{P}(0) \} \). 
The numerical integration is carried out outwards starting from a small but finite radius \(r_0 > 0\). 
The lapse is initially chosen to be \(N(0) = 1\) and is rescaled after the integration so that it approaches $N=1$ at spatial infinity (this is possible by rescaling the time coordinate by a constant factor).
Given a central pressure \(\tilde{P}(0)\), the integration constant \(\varphi_0\) is fixed by a shooting method so that $\varphi$ asymptotes to zero at infinity.
The location $r_\star$ of the surface of the star is determined by $\tilde{P}\left(r_\star\right) = 0$.
The baryon mass in the Jordan frame is calculated as
\begin{align}
    \tilde{M}_b \equiv \int d^{3}\tilde{x} \sqrt{-\tilde{g}} \tilde{\rho}_0 \tilde{u}^{0}
\end{align}
whereas the
scalar charge is calculated as~\cite{Damour:1992we,Palenzuela:2013hsa, Bezares:2021yek}
\begin{align}
    \alpha_c \equiv \sqrt{\dfrac{4 \pi}{G}} \dfrac{\varphi_1}{M_\infty}~,
\end{align}
where \(\varphi_1\) appears in the asymptotic expansion \(\varphi(r) = \varphi_\infty + \varphi_1 r^{-1} + \mathcal{O}(r^{-2})\), and \(M_\infty\) is the gravitational mass in the Einstein frame appearing in the asymptotic expansion \(N^{2}(r)= 1 - 2 G M_\infty r^{-1} + \mathcal{O}(r^{-2})\).

%%%%%%%%%%%%%%%%%%%%%%%%%%%%%%%%%%%%%%%%%%%%%%%%%%%%%%%%%%%%%%%%
%%%%%%%%%%%%%%%%%%%%%%%%%%%%%%%%%%%%%%%%%%%%%%%%%%%%%%%%%%%%%%%%
%%%%%%%%%%%%%%%%%%%%%%%%%%%%%%%%%%%%%%%%%%%%%%%%%%%%%%%%%%%%%%%%

\subsection{Simplified analytic model} \label{sec: simplified analytic model}

We complement our numerical analysis by introducing an analytical toy model for the scalar equation of motion, which, we anticipate, will allow us to reach values for the strong coupling scale relevant for dark energy, i.e. \(\Lambda_{\text{DE}} \sim 1 \, \mathrm{meV}\) (these values are hard to simulate numerically due to the hierarchy of scales involved~\cite{terHaar:2020xxb, Bezares:2021yek}).
Moreover, our model will allow us to cross check and interpret certain features of the numerical scalar profiles presented in the next Section.
This model can be applied to all matter couplings in Eq.~\eqref{eq: Phi definition} but $\varphi^2$, 
because
in the latter case
the source of the scalar equation is a function of \(\varphi\). 
Therefore, in that case the solution can only be obtained numerically as described in Sec.~\ref{sec: numerical integration}.

In more detail, we approximate the spacetime to be Minkoswki and consider a star described by the Tolman VII EOS~\cite{Tolman:1939jz},
\begin{align} \label{eq: Tolman VII EOS}
    \tilde{\rho} \left(r\right) \equiv \tilde{\rho}_c \left[1-\left(\dfrac{r}{r_\star}\right)^{2}\right] ~,
\end{align}
where \(r_\star\) is the star surface location and \(\tilde{\rho}_c\) is a parameter specifying the central energy density.
We then define\footnote{
Since we are looking at static solutions in spherical symmetry, the decomposition of $\mathcal{K}\left(\varphi, X\right) \varphi_\mu$ in  the gradient part only is complete.
}
\begin{align} \label{eq: Psi gradient definition}
    \partial_\mu \chi \equiv \mathcal{K} \left(\varphi, X\right) \varphi_\mu ~,
\end{align}
so that the scalar equation becomes
\begin{align} \label{eq: flat space scalar equation}
    \nabla^2 \chi = - \hat{\alpha} \tilde{T}~,
\end{align}
where \(\hat{\alpha} \equiv \alpha_1 / (4 \Phi^2 M_\text{Pl})\).
Approximating \(\tilde{T} \approx - \tilde{\rho}\) and \(\Phi \approx 1\), one can solve Eq.~\eqref{eq: flat space scalar equation} by making use of the Green function \(\mathcal{G}(\boldsymbol{x}, \boldsymbol{x}') \equiv -1/(4 \pi \lvert \boldsymbol{x}- \boldsymbol{x'}\rvert)\) and obtain
\begin{align} \label{eq: Sol Psi}
    \chi' = \begin{cases}
    \hat{\alpha} \tilde{\rho}_c \left(\dfrac{r}{3} - \dfrac{r^3}{5r^2_\star}\right), & r \leq r_\star~, \\
     \dfrac{2 \hat{\alpha} \tilde{\rho}_c r_\star^3}{15 r^2}, & r_\star < r ~.
    \end{cases}
\end{align}
Finally, we invert Eq.~\eqref{eq: Psi gradient definition} to obtain \(\varphi'\).

As an illustration, and for the sake of simplicity of the analytic expressions, we present here the case
where $\beta=\lambda_1=0$ and $\lambda_2<0$ in Eqs.~\eqref{eq: Kes definition}-\eqref{eq: Phi definition}.
The extension to more general cases is straightforward. 
For this case, 
$\mathcal{K}$ can be 
% easily 
written as 
\begin{align} 
\label{eq: def of cal K for quadratic example}
    \mathcal{K} = -\dfrac{1}{2}\left(1 + \hat{\lambda} \tilde{T} X \right)~,
\end{align}
where
\( \hat{\lambda} \equiv 2 \lambda_2 / (\Phi^2 \Lambda^8) \), and
the solution for \(\varphi'\) is given by the analytic inversion of Eq.~\eqref{eq: Psi gradient definition},
which corresponds to the real root \(y \equiv \varphi'^2\) of the third-order polynomial 
\begin{align} \label{eq: cubic polynomial}
    \left(1 -\hat{\lambda} \tilde{\rho} y\right)^2 y - 4\chi'^2 = 0~.
\end{align}
The full solution
\textcolor{red}{
}
is not particularly illuminating.
However, it is 
instructive to examine separately the linear and non-linear regimes. 
The linear solution is relevant 
near the center of the star ($\varphi' \propto r$) and in the exterior ($\varphi' \propto r^{-2}$). In the not-so-deep interior of the star, however, the nonlinear term dominates and, if the density gradients are small, gives \(\varphi' \propto r^{1/3}\). 

\section{Results} \label{sec: results}

In this Section we illustrate neutron star solutions for the different terms in the conformal coupling \eqref{eq: Phi definition}.
First, in Sec.~\ref{sec: screened scalarization}, we consider the effect of a quadratic coupling in $\varphi$ on screened solutions in $k$-essence, and we investigate whether scalarization can take place even in this scenario.
Then, in Sec.~\ref{sec: quadratic kinetic coupling }, we study the effect of \(X\)-dependent couplings on 
FJBD theory and
finally, in Sec.~\ref{sec: last example }, we show the effect of the same kinetic couplings on $k$-essence theories that have screening. 

\subsection{Couplings to matter dependent on \(\varphi\)} \label{sec: screened scalarization}

\begin{figure}[h!]
 \centering
\includegraphics[width=3.4 in]{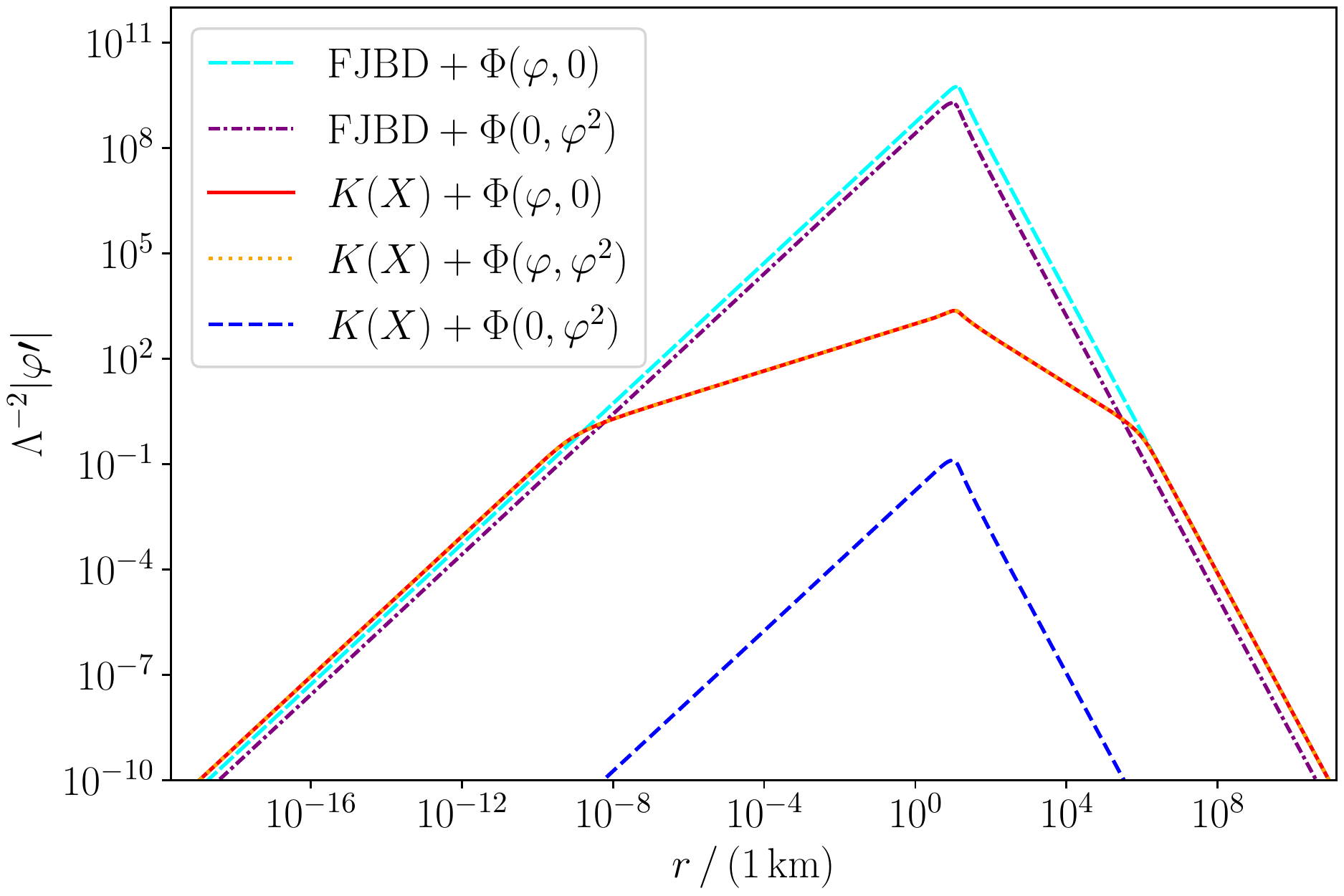}
\caption{\emph{Scalar profile for \(\varphi\)-dependent matter couplings.} 
Dashed cyan: Standard FJBD solution with parameters \((\beta, \alpha_1, \alpha_2) = (0, 1, 0)\).
Dot dashed purple:
DEF model with parameters
\((\beta, \alpha_1, \alpha_2) = (0, 0, 9/4)\).
Solid red:
\(k\)-essence with 
\((\beta, \alpha_1, \alpha_2) = (-1, 1, 0)\).
Dotted orange:
\(k\)-essence with
\((\beta, \alpha_1, \alpha_2) = (-1, 1, 9/4)\).
Dashed blue:
\(k\)-essence with
\((\beta, \alpha_1, \alpha_2) = (-1, 0, 9/4)\).
We choose stars with \(\tilde{M}_b = 1.87 \, M_\odot\),
near the top of the \(\beta = \alpha_1 = 0\) charge-mass curve. 
In all cases we use
\(\Lambda = 1 \, \mathrm{keV}\).
}
\label{fig: scalar profile keV}
\end{figure}

We restrict to the part of Eq.~\eqref{eq: Phi definition} depending only on the scalar field and not on its derivatives, i.e. $\lambda_1=\lambda_2=0$.
In Fig.~\ref{fig: scalar profile keV}, we illustrate this case with an example of a generic solution with \(| \alpha_{1, 2}| \sim \mathcal{O}(1)\). 
We show 
the scalar gradient profile for FJBD theory with a linear coupling to matter (light-blue dashed) and with a quadratic one (purple dotdashed). 
The former is only shown for comparison as this case is ruled out by solar system constraints.
The latter is
usually referred to as the Damour-Esposito-Far{\`e}se (DEF) model, which leads to scalarized neutron stars.
We also show, as a reference, the standard screened solution of $k$-essence \cite{terHaar:2020xxb}, where only the linear coupling is present (red solid line).
In this case, we observe the presence of a screening region (i.e. a change in slope which results in a suppression of the scalar force) between the first and last knee of the scalar gradient, with the location of the last knee (counting from the left) corresponding to the screening radius.

When we turn on a quadratic coupling in $\varphi$ (orange dotted line), we observe no apparent difference with respect to the screened solution described above (red solid line). 
This is not surprising since higher-order couplings in $\varphi$ are suppressed by the corresponding power of \(M_\text{Pl}\), which makes them irrelevant unless one hits the Planck scale.
Note that this would be the case even in FJBD theory, unless one were to set $\alpha_1=0$ like in the example above, or to a very small value compatible with 
solar system constraints. In FJBD theory, the latter do indeed bound $|\alpha_1| < 5 \times 10^{-3}$, thus making the quadratic coupling dominant. 

Although 
there is no need for a small $\alpha_1$ coupling in $k$-essence
(due to screening), 
it is nevertheless intriguing to investigate such possibility and address the question of whether scalarization can still take place. Setting $\alpha_1=0$ and leaving only the quadratic coupling active
(in analogy to the DEF model), 
we find
that the magnitude of the scalar gradient is uniformly suppressed (blue dashed line). The scalar field remains always in the linear regime [i.e. \(\varphi' / \Lambda^2 \lesssim \mathcal{O}(1)\)] and, as a consequence,  
when imposing a vanishing scalar field at infinity, the maximum of $\varphi'$ can never exceed the value of $\Lambda^2$. 

We empirically observe
a uniform suppression that scales roughly as \(\varphi' \propto \Lambda^2 \), which in turn leads to a suppression of the scalar charge as \(\alpha_c \propto \Lambda^2\) with respect to the DEF model. 
Therefore, we can conclude that, even if the non-linear regime of $k$-essence never kicks in when only a quadratic coupling with matter is present, there is still an overall suppression of the scalar force which depends on $\Lambda$. 
This implies a suppression of the scalar charge, which makes  scalarization impossible in the presence of $k$-essence.

\begin{figure}[h!]
 \centering
\includegraphics[width=3.4 in]{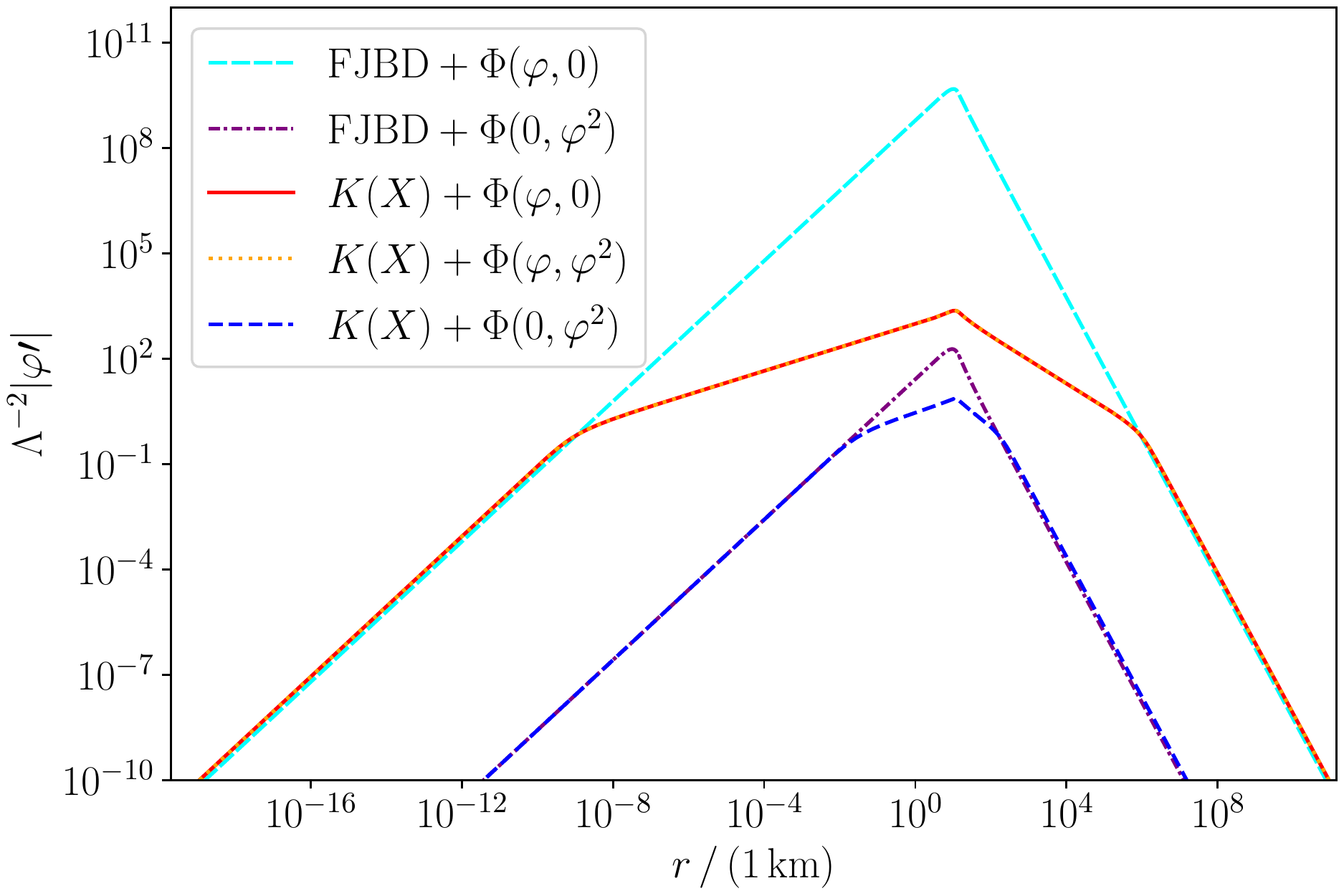}
\caption{\emph{Scalar profile for \(\varphi\)-dependent matter couplings with equal \(\varphi(0)\).} 
Same cases as in Fig.~\ref{fig: scalar profile keV}. Here, however, we fix the boundary condition at the center to the same constant value \(\varphi(0) / \Lambda = 1.709 \times 10^{-3}\). %
}
\label{fig: scalar profile keV Phi0 Test}
\end{figure}

Let us now further investigate the origin of this uniform suppression
and how it is affected by 
the boundary conditions. In Fig.~\ref{fig: scalar profile keV Phi0 Test}, we plot the same cases as in Fig.~\ref{fig: scalar profile keV}, but instead of imposing a vanishing scalar field at infinity, we fix the central scalar field to the same constant value in all cases. We choose the value that was previously obtained for the \(k\)-essence profile with linear matter coupling (red solid line).
Since the perturbative expansion around \(r = 0\) is the same in FJBD theory and \(k\)-essence, 
we obtain two classes of solutions 
corresponding to the different matter couplings. 
Despite the same central scalar field, the value of the scalar field at infinity is now different for the two classes (due to the different field equations), and in the quadratic coupling case it does not vanish.
This translates into larger scalar gradients, which make the $k$-essence solution with quadratic coupling enter the non-linear regime. Therefore, in order to excite $k$-essence non-linearities even with a quadratic coupling to matter, one would need a non-trivial scalar field
boundary condition at infinity  enhancing the scalar gradient.

%

%%%%%%%%%%%%%%%%%%%%%%%%%%%%%%%%%%%%%%%%%%%%%%%%%%%%%%%%%%%%%%%%
%%%%%%%%%%%%%%%%%%%%%%%%%%%%%%%%%%%%%%%%%%%%%%%%%%%%%%%%%%%%%%%%
%%%%%%%%%%%%%%%%%%%%%%%%%%%%%%%%%%%%%%%%%%%%%%%%%%%%%%%%%%%%%%%%

\subsection{FJBD and kinetic coupling with matter} \label{sec: quadratic kinetic coupling }

We now study FJBD theory with
a coupling to matter  depending on
the kinetic term \(X\).
Note that, 
in order to break the shift symmetry of the theory and avoid no-hair theorems 
(see e.g.~Ref.~\cite{Lehebel:2017fag}), 
we also need to maintain a dependence on \(\varphi\), which we assume to be linear.

\begin{figure}[h!]
 \centering
\includegraphics[width=3.4 in]{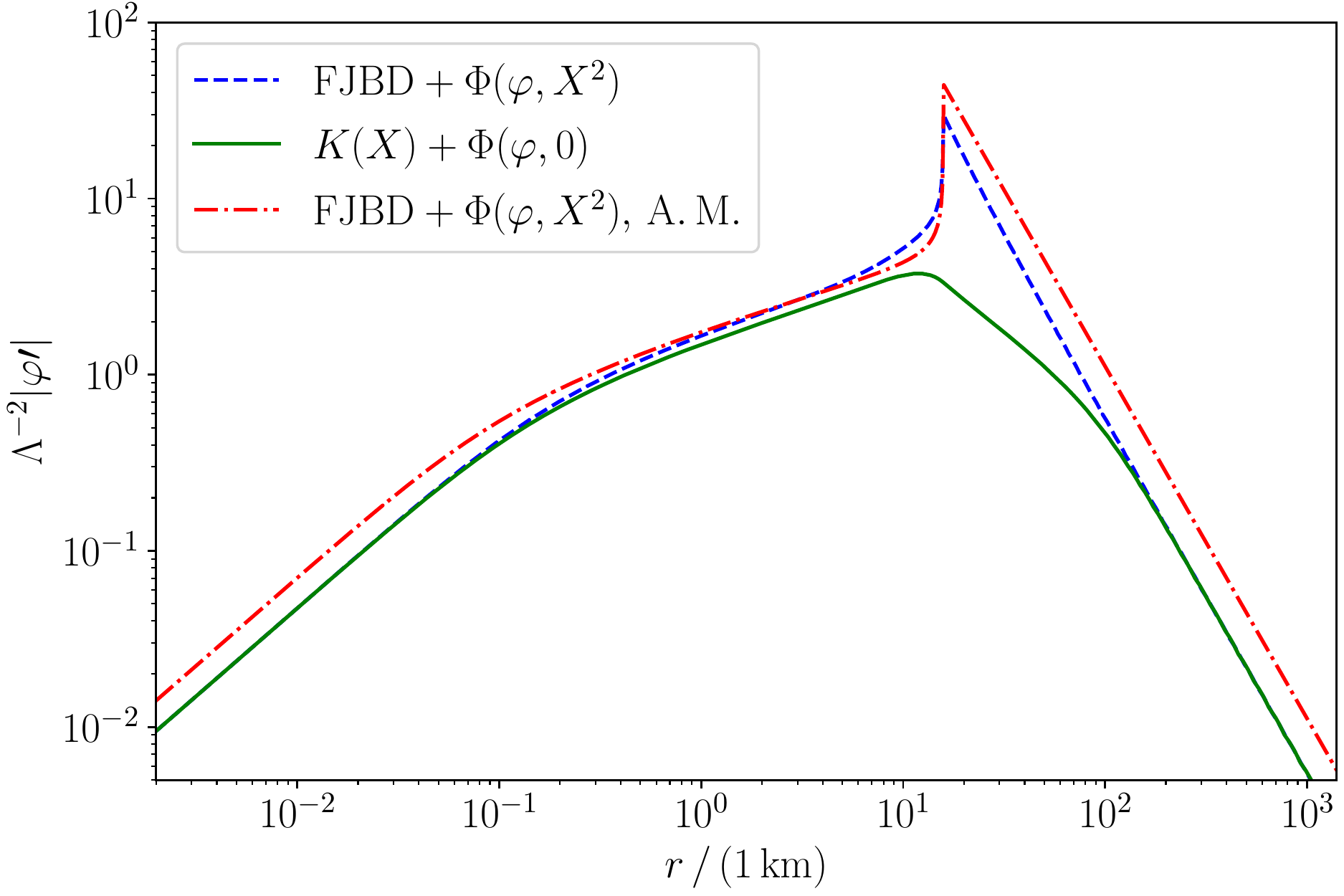}
\caption{\emph{FJBD theory with \(X^2\)-coupling to matter.} 
Dashed blue: scalar gradient profile for \((\beta, \alpha_1, \lambda_1, \lambda_2) = (0, 10^{-3}, 0, -1)\)
and
\(\Lambda =  10.3 \, \mathrm{MeV}\).
Solid green: kinetic screening in \(k\)-essence with parameters \((\beta, \alpha_1, \lambda_1, \lambda_2) = (-1, 10^{-3}, 0, 0)\) -- for the purposes of comparison  we have used an energy scale \(\Lambda = 0.38 \, \mathrm{MeV}\).
Red dash dot: scalar gradient profile obtained with the analytic model (A.~M.) with parameters
\((\beta, \alpha_1, \lambda_1, \lambda_2) = (0, 10^{-3}, 0, -1)\)
and \(\Lambda =  10.3 \, \mathrm{MeV}\).
We choose \(\tilde{\rho}_c = \tilde{\rho}\left(0\right)\) and \(r_\star\) 
as given by the integration in \(k\)-essence. 
}
\label{fig: example1 lambda negative}
\end{figure}

As an illustrative example, and to validate our analytic model, we begin by considering the case of only including the \(X^2\) term in \(\Phi\).
In Fig.~\ref{fig: example1 lambda negative},
we show the numerical solution for the scalar gradient profile (blue dashed line) for
\(\lambda_2 = -1\) and \(\Lambda \approx 10.3 \, \mathrm{MeV}\). 
We observe a behavior similar to kinetic screening in the interior of the star, whereas in the exterior the scalar behaves linearly.
For comparison, we show the solution in \(k\)-essence (green solid line) for $\beta=-1$ and \(\Lambda = 0.38 \, \mathrm{MeV}\), which comes from the matter redressing of the above energy scale, i.e.
\(\Lambda^4_{\text{eff}} = \Lambda^8 / [2 \tilde{\rho}\left(0\right)]\).
In more detail, this effect can be 
understood by noting that the quadratic term in the conformal coupling and in \(k\)-essence enter in the same way in the scalar equation \eqref{eq: scalar equation}, 
namely 
\(2\mathcal{K}+1 \propto \lambda_2  \tilde{\rho} X / \Lambda^{8} \) 
and 
\(2\mathcal{K}+1 \propto \beta X / \Lambda^{4} \),
respectively.
The presence of \(\tilde{\rho}\) in the first case is responsible for the different energy scale at which screening takes place.
Moreover, since this effect is proportional to the density distribution, 
it fades away as the star surface is approached (as  \(\tilde{\rho} \to 0\)), and thus it is not present in the exterior. 

In Fig.~\ref{fig: example1 lambda negative} we also note the appearance of a ``cusp'' at the star surface, \(r_\star \approx 14 \, \mathrm{km}\), 
which arises from the connection between the non-linear (interior) and linear (exterior) behavior.
Despite this feature, 
we have checked that 
all quantities 
(including the curvature invariants \(R\),  \(R_{\mu\nu}R^{\mu\nu}\), and \(R_{\mu\nu\rho\sigma}R^{\mu\nu\rho\sigma}\)) remain regular at \(r_\star\).
From a practical point of view, 
numerically integrating this cusp is challenging, as one must make sure that the scalar gradient 
is resolved, and for lower values of \(\Lambda\) this becomes increasingly difficult to achieve.

In order to more easily capture the behavior at the star surface, and to cross-check our numerical scalar profiles, we resort to the analytical model introduced in Sec.~\ref{sec: simplified analytic model}.
In Fig.~\ref{fig: example1 lambda negative}, we show the scalar gradient profile (red dot dashed line) generated with the analytic model, where the parameters of the equation of state~\eqref{eq: Tolman VII EOS} are chosen to closely match the stellar radius and central density of the numerical integration profile.
We observe that the scalar gradient predicted by this model is qualitatively similar to the numerical one (blue dashed line), with differences mainly due to the details of the EOS.
Moreover, in the analytic model we can perform an expansion near the surface of the star to confirm that the scalar gradient approaches a finite value at the cusp. Expanding in \(\Delta r \equiv r - r_\star\), as \(\Delta r \to 0^{-}\), we obtain
\begin{multline} \label{eq: real cubic solution expansion 2}
    \pm \varphi' = \pm \sqrt{y} = \dfrac{4 }{15}\hat{\alpha} \tilde{\rho}_c r_\star \\
    + \dfrac{8}{3375} \hat{\alpha}  \tilde{\rho}_c \left(225 + 16  \hat{\lambda}\hat{\alpha}^2 \tilde{\rho}_c^3 r_\star^2 \right)\Delta r \\
    + \mathcal{O}\left[\left(\Delta r\right)^2\right] ~ ,    
\end{multline}
where the overall sign can be fixed by comparing with the interior solution.

\begin{figure}[h!]
 \centering
\includegraphics[width=3.4 in]{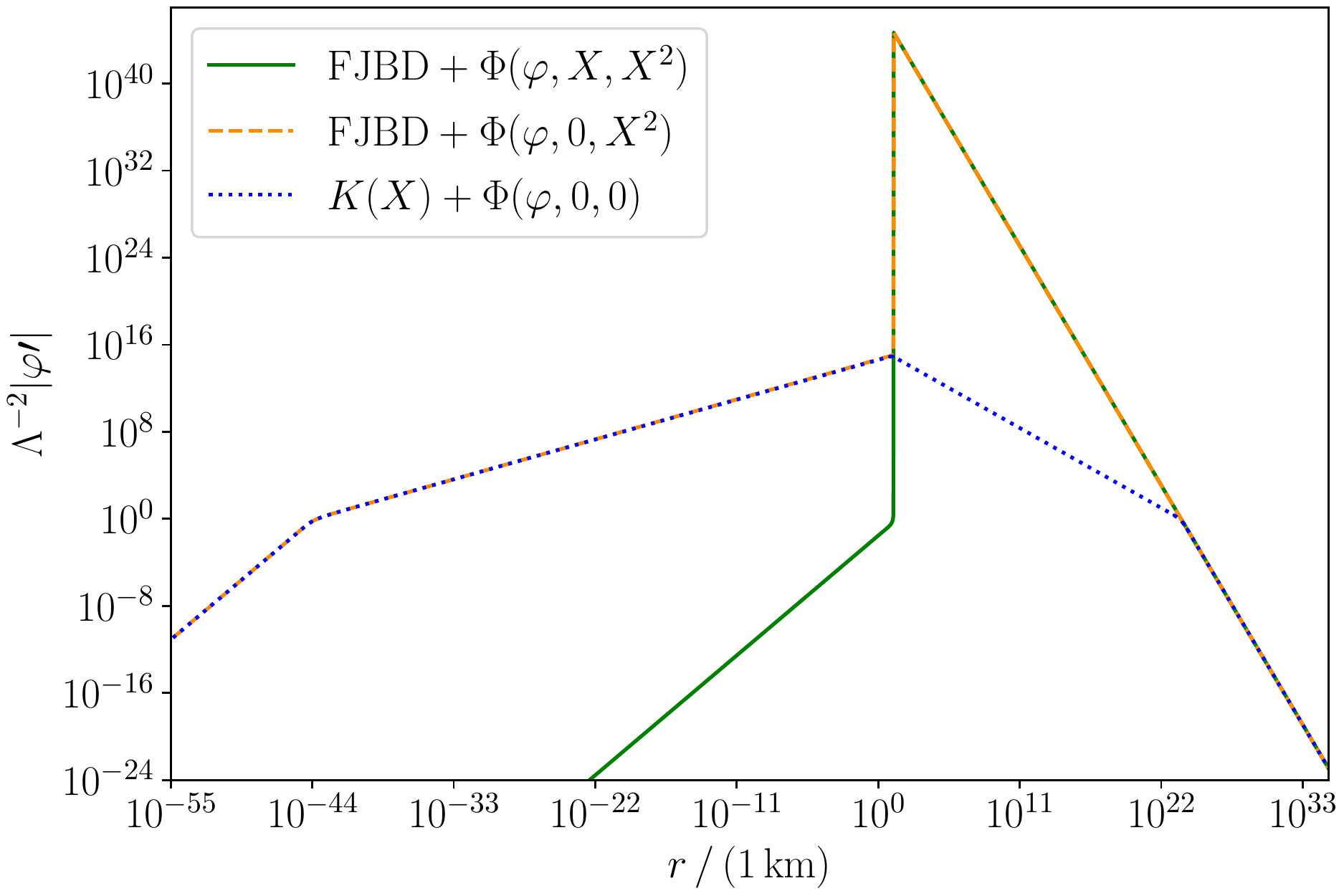}
\caption{\emph{FJBD theory with generic kinetic coupling to matter at dark energy scales.} 
Solid green: scalar gradient profile for
\((\beta, \alpha_1, \lambda_1, \lambda_2) = (0, 1, -1, -1)\).
Dashed orange: scalar gradient profile for
\((\beta, \alpha_1, \lambda_1, \lambda_2) = (0, 10^{-3}, 0, -1)\).
In both solutions, the energy scale is \(\Lambda = 1 \, \mathrm{meV}\).
Dotted blue: for comparison, the \(k\)-essence screened solution with \((\beta, \alpha_1, \lambda_1, \lambda_2) = (-1, 10^{-3}, 0, 0)\) and \(\Lambda = 3.6 \times 10^{-12}\, \mathrm{meV}\). 
The vertical axis is rescaled with respect to the latter value for \(\Lambda\)
and,
in all cases, 
we use the analytic model. 
}
\label{fig: example1 lambda negative 2}
\end{figure}

Having validated our analytic model, we can now employ it to study scales 
relevant for dark energy.
In Fig.~\ref{fig: example1 lambda negative 2}, we show the scalar gradient profile (green solid line) for \(\lambda_1 = \lambda_2 = -1\) in Eq.~\eqref{eq: Phi definition} and energy scale \(\Lambda \sim \Lambda_\text{DE} \sim 1 \, \mathrm{meV}\).
For comparison, we also plot (orange dashed line) the case studied above (i.e. $\lambda_1 = 0$ and $\lambda_2 = -1$), and (blue dotted line) the profile in \(k\)-essence (\(\beta = -1\)) with an equivalent strong-coupling scale \(\Lambda \approx 3 \times 10^{-12} \, \mathrm{meV}\).
In particular, we observe that the scalar field is always in the linear regime, with a different magnitude of the gradient inside and outside the star.
The value inside is greatly suppressed, as compared to the value outside, due to the redressing of the linear term in the scalar equation inside matter.

To conclude this Section,
and for completeness, 
we briefly summarize the behavior for the positive signs of the kinetic couplings in Eq.~\eqref{eq: Phi definition}.
For 
\(\lambda_1  > 0\), 
there is a location inside the star where \(\mathcal{K} = 0\) and where the scalar gradient diverges and changes sign. 
This can be seen from the analog of Eq.~\eqref{eq: def of cal K for quadratic example} which is given by 
\(2\mathcal{K} = -1 + \hat{\lambda}_1 \tilde{\rho}(r)\), 
where
\(\hat{\lambda}_1 \equiv \lambda_1 / (2 \Phi^2 \Lambda^4)\).
In this case, after substituting our approximation for the EOS [Eq.~\ref{eq: Tolman VII EOS}],
one can show that if \(\lambda_1 > 0\), 
the scalar becomes singular when 
\(\mathcal{K}\)
vanishes. 
The condition for vanishing \(\mathcal{K}\) is 
\(r_{p}^2 / r_{\star}^2 = 1-1/(\hat{\lambda}_1 \tilde{\rho}_c) \). 
Therefore, this is a pathological solution. 
Finally, for  
\(\lambda_1  = 0\) and \(\lambda_2 > 0\), 
the solutions become pathological
as
the root of Eq.~\eqref{eq: cubic polynomial}, with regular boundary condition at the origin (\(\varphi'(0) = 0\)), is not real in the entire radial domain. 
Indeed, when plotting this quantity we notice that it acquires an imaginary part for radii at which~\footnote{
For the example of Sec.~\ref{sec: simplified analytic model}, the root of the cubic polynomial~\eqref{eq: cubic polynomial} that has the correct boundary conditions at the origin (\(\varphi'(0)=0\)) 
is
\(
    y 
    =
    -(q^{1/3}-1)^2/(3 \hat{\lambda} \tilde{\rho}  q^{1/3} ) 
\)
where
\(
    q 
    \equiv 
    6 \chi ' \left( [-3\hat{\lambda}  \tilde{\rho}  \mathcal{D}]^{1/2}-9
   \hat{\lambda}  \tilde{\rho}  \chi '\right)+1 
\),
and
\(
    \mathcal{D}
    \equiv
    1-27 \hat{\lambda}  \tilde{\rho}  \chi'^2   
\).
}
\(\mathcal{D} \equiv 1-27 \hat{\lambda}  \tilde{\rho}(r)  {\chi'}^2 (r) < 0 \).
This situation can be easily achieved for parameters of Fig.~\ref{fig: example1 lambda negative 2} (but with \(\lambda_2 = 1\)) and \(\Lambda \lesssim 100 \, \mathrm{MeV}\).
For the scales relevant for dark energy [\(\mathcal{O}(1)\,\mathrm{meV}\)] this is  clearly problematic.
%

%
%%%%%%%%%%%%%%%%%%%%%%%%%%%%%%%%%%%%%%%%%%%%%%%%%%%%%%%%%%%%%%%%
%%%%%%%%%%%%%%%%%%%%%%%%%%%%%%%%%%%%%%%%%%%%%%%%%%%%%%%%%%%%%%%%
%%%%%%%%%%%%%%%%%%%%%%%%%%%%%%%%%%%%%%%%%%%%%%%%%%%%%%%%%%%%%%%%

\subsection{$k$-essence and kinetic coupling with matter} \label{sec: last example }

In this Section,
we rely on our analytic model to study the effects of kinetic coupling with matter in $k$-essence, for dark energy scales.

\begin{figure}[]
 \centering
\includegraphics[width=3.4 in]{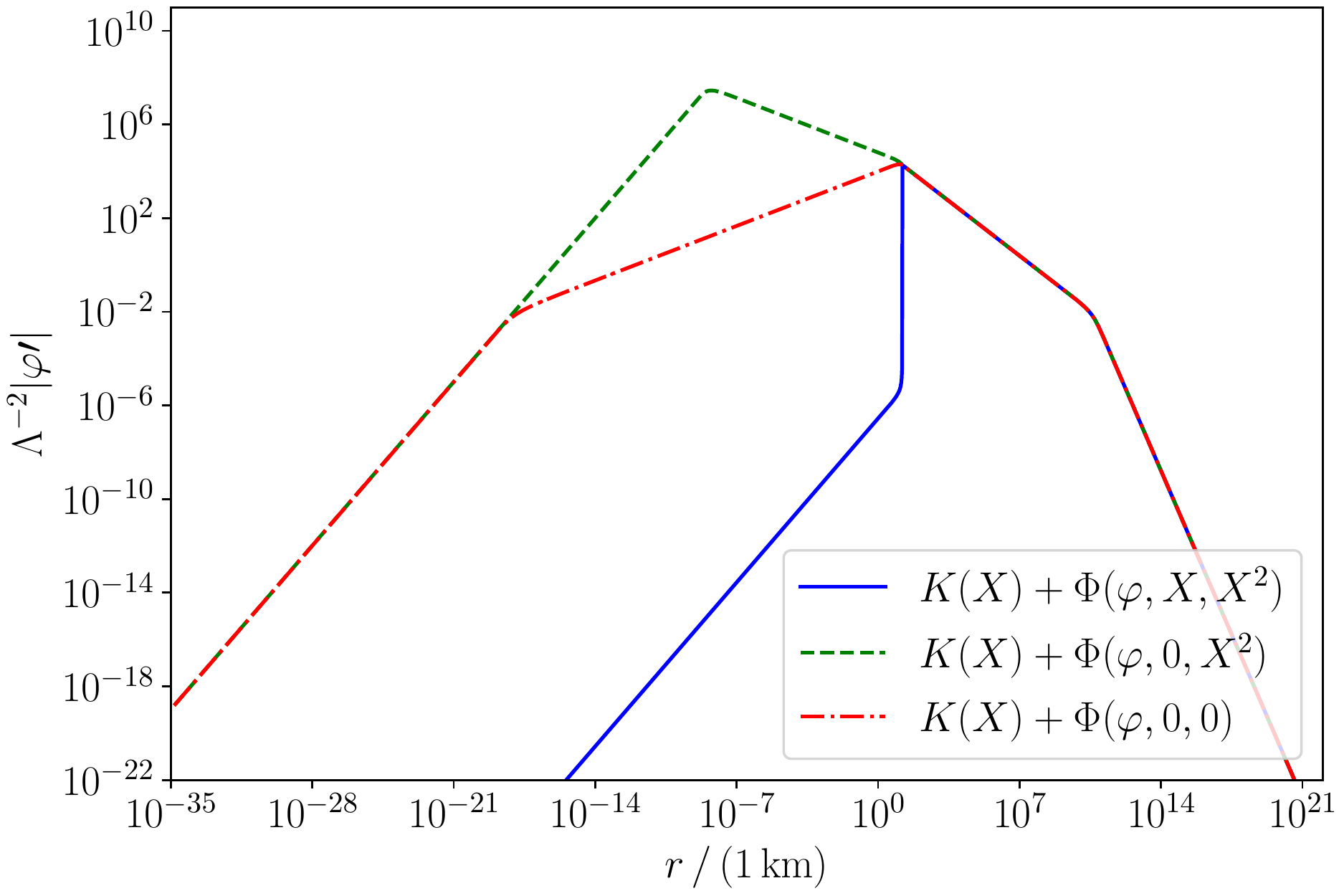}
\caption{\emph{\(k\)-essence and kinetic coupling with matter.} 
Solid blue:
With parameters \((\beta, \alpha_1, \lambda_1, \lambda_2) = (-1, 10^{-3}, -1, 1)\).
Dashed green:
With parameters \((\beta, \alpha_1, \lambda_1, \lambda_2) = (-1, 10^{-3}, 0, 1)\) and  \(\Lambda = 1 \, \mathrm{meV} \) for the function \(K(X)\), however, we choose a different energy scale
\(\Lambda \approx 0.52 \, \mathrm{keV}\) in \(\Phi\) for illustrative purposes, such that the cancellation effect is maximized.
Dash dot red: \((\beta, \alpha_1, \lambda_1, \lambda_2) = (-1, 10^{-3}, 0, 0)\) and 
\(\Lambda = 1 \, \mathrm{meV}\). 
The parameters used for the equation of state \(\{\tilde{\rho}_c, r_\star\}\) are taken to be the same as in Fig.~\ref{fig: example1 lambda negative}. }
\label{fig: simplified linear model kes antiscreening}
\end{figure}

For \(\lambda_1 < 0\), and regardless of the sign of $\lambda_2$, we observe an enhancement of the scalar suppression inside the star, in line with the previous discussion. 
In fact, as shown in Fig.~\ref{fig: simplified linear model kes antiscreening} (solid blue), inside the star the scalar field remains always in the linear regime with a gradient uniformly suppressed as in
the previous Section, 
while kinetic screening remains unchanged in the exterior of the star.

More interesting
phenomenology 
occurs
when $\lambda_1=0$ and \(\lambda_2 > 0\).
Indeed, in this case, 
the scalar gradient profile features a 
\emph{weakening}
of the kinetic screening inside the star, due to the partial cancellation of the non-linear terms coming from $k$-essence and from the matter coupling. 
In Fig.~\ref{fig: simplified linear model kes antiscreening}, we show this effect (green dashed line). 
For demonstrative purposes, and in order to maximize this effect, we chose a different energy scale for the matter coupling given by \(\Lambda_{\text{eff}}^8\approx 2 \Lambda^4 \tilde{\rho}_c \approx (0.52 \, \mathrm{keV})^8\). 
This effect is similar, although different in nature, to the partial breaking of the Vainshtein screening inside matter for beyond Horndeski \cite{Kobayashi:2014ida, Babichev:2016jom} and DHOST theories \cite{Crisostomi:2017lbg, Langlois:2017dyl, Dima:2017pwp, Crisostomi:2019yfo}.
Finally, for comparison, we also show (red dot dashed) the profile for a screened star in \(k\)-essence without kinetic coupling to matter.

%

%%%%%%%%%%%%%%%%%%%%%%%%%%%%%%%%%%%%%%%%%%%%%%%%%%%%%%%%%%%%%%%%
%%%%%%%%%%%%%%%%%%%%%%%%%%%%%%%%%%%%%%%%%%%%%%%%%%%%%%%%%%%%%%%%
%%%%%%%%%%%%%%%%%%%%%%%%%%%%%%%%%%%%%%%%%%%%%%%%%%%%%%%%%%%%%%%%

\section{Conclusions} \label{sec: Conclusion}

In this paper, we have investigated the effect of  general matter couplings for \(k\)-essence scalar-tensor theories featuring kinetic screening of local scales.
We summarize our findings in Table~\ref{tab: summary table}.
% 
%
% \begin{widetext}
\begin{table}
\caption{\label{tab: summary table}
Summary of the cases explored.
}
\begin{ruledtabular}
\begin{tabular}{c|c|c}
\textrm{Case} &  \textrm{Matter coupling } & \textrm{Effect(s)} \\
 & \((\alpha_1, \alpha_2, \lambda_1, \lambda_2)\) & \\
\colrule
%\hline
\hline
FJBD\(+K(X)\) & \((+, 0, 0, 0)\) & Screening \\
 (Fig.~\ref{fig: scalar profile keV}) &  &  \\
\hline 
DEF &  \((0, +, 0, 0)\) & Scalarization \\
(Fig.~\ref{fig: scalar profile keV}) &  &  \\
\hline 
DEF\(+ K(X)\)  & \((0, +, 0, 0)\) & No screening \\
(Case I, Fig.~\ref{fig: scalar profile keV}) &  & No scalarization\\
\hline 
DEF\(+ K(X)\)  &  \((+, +, 0, 0)\) & Screening \\
(Case II, Fig.~\ref{fig: scalar profile keV}) &  & \\
\hline 
FJBD\(+\Phi(\varphi,X)\) &  \((+, 0, -, 0)\) & Linear suppr.  \\
 (Not shown) &  & (interior)  \\
\hline  
FJBD\(+\Phi(\varphi,X)\) & \((+, 0, +, 0)\) & Pathological.  \\
 (Not shown) &  & (interior)  \\ 
\hline  
FJBD\(+\Phi(\varphi, X^2)\) & \((+, 0, 0, -)\) & Screening  \\
 (Fig.~\ref{fig: example1 lambda negative 2}) & & (interior)  \\
\hline  
FJBD\(+\Phi(\varphi, X^2)\) & \((+, 0, 0, +)\) & Pathological.  \\
 (Not shown) & & (interior)  \\ 
\hline  
FJBD\(+\Phi(\varphi,X,X^2)\) & \((+, 0, -, -)\) & Linear suppr.  \\ 
% &  &  & scalar suppr.\\
 (Fig.~\ref{fig: example1 lambda negative 2})  & & (interior)\\
\hline   
FJBD\(+K(X)\) & \((+, 0, 0, +)\) & Anti-screening \\ 
\quad \(+\Phi(\varphi,X^2)\)   & & (interior)\\
 (Fig.~\ref{fig: example1 lambda negative}) & & \\
\hline 
FJBD\(+K(X)\) &  \((+, 0, -, -)\) & Linear suppr.  \\ 
\quad \(+\Phi(\varphi,X,X^2)\)   &  & (interior)\\
  (Fig.~\ref{fig: example1 lambda negative}) &  & Screening (exterior) \\
\end{tabular}
\end{ruledtabular}
\end{table}
% \end{widetext}

In Refs.~\cite{terHaar:2020xxb, Bezares:2021yek, Bezares:2021dma}, the standard linear coupling in \(\varphi\) was considered in 
studies of neutron star oscillations and mergers. 
We find here that this choice is robust 
against more general matter couplings, at least for the static configurations chosen as initial data. Indeed,  we have explicitly shown that the inclusion of a quadratic term in \(\varphi\) does not affect significantly the solution.
Moreover, even in a configuration \emph{\`a la} DEF (i.e. with a matter coupling including only a quadratic term in \(\varphi\)), scalarization
becomes negligible
as
the scalar gradient profile becomes even more suppressed (although always in the linear regime).

We have also explored couplings to matter depending explicitly on the kinetic term of the scalar field. Such terms arise from $X$-dependent conformal transformations of viable DHOST theories in the Jordan frame. 
We have identified healthy sectors of the parameter space, where the suppression of the scalar gradient inside the star is enhanced, while the theory is still in the linear regime. 
This in turn produces a sharp transition between the linear and non-linear regimes at the star surface, which is however devoid of pathologies.

Finally, 
interesting phenomenology arises when only a quadratic kinetic term is present in the matter coupling (together with the usual linear term in $\varphi$). 
In this case, 
there is a weakening of the kinetic screening inside 
stars, due to the partial cancellation of the non-linear terms coming from $k$-essence and from the matter coupling. 

This new partial breaking of the screening mechanism inside matter has several consequences for stellar astrophysics, which can be used to test the model. For example, the Chandrasekhar mass limit \cite{Jain:2015edg}, the burning process in brown and red dwarfs \cite{Sakstein:2015zoa} and the mass-radius relation of white dwarfs \cite{Saltas:2018mxc} are all quantities sensitive to the gravitational force inside bodies. These observables have been used to place constraints on models which, like ours, break the screening inside matter. We plan to explore these effects and use them to constrain kinetic couplings in future work.

\begin{acknowledgements}

We would like to thank Nicola Franchini and Lotte ter Haar for insightful discussions. 
G.~L.~thanks the Perimeter Institute for Theoretical Physics for hospitality while this work was being completed. 
All authors acknowledge financial support provided under the European Union's H2020 ERC Consolidator Grant ``GRavity from Astrophysical to Microscopic Scales'' grant agreement no. GRAMS-815673. 
This work was supported by the EU Horizon 2020 Research and
Innovation Programme under the Marie Sklodowska-Curie Grant Agreement
No. 101007855.

\end{acknowledgements}

%%%%%%%%%%%%%%%%%%%%%%%%%%%%%%%%%%%%%%%%%%%%%%%%%%%%%%%%%%%%%%%%
%%%%%%%%%%%%%%%%%%%%%%%%%%%%%%%%%%%%%%%%%%%%%%%%%%%%%%%%%%%%%%%%
%%%%%%%%%%%%%%%%%%%%%%%%%%%%%%%%%%%%%%%%%%%%%%%%%%%%%%%%%%%%%%%%

\appendix

\section{Derivation of the scalar and matter equations} \label{sec: app EOMs}

Variation of the action with respect to \(\varphi\) and the Bianchi identity applied to Eq.~\eqref{eq: Einstein equations} give
\begin{align}
\nabla_\mu \left[\mathcal{K} \varphi^\mu \right]
= - \dfrac{\tilde{T}}{4 \Phi^3}  \Phi_\varphi ~, \label{eq: scalar equation 2}\\
\label{eq: matter equation 2}
\nabla_{\mu} \left[T^{\mu \nu} + T^{(\varphi)\, \mu \nu}\right]
=  0 ~,
\end{align}
respectively,
where
\begin{align} \label{eq: definition Q 2}
    \mathcal{K}\equiv K_X - \dfrac{\tilde{T}}{2 \Phi^3}  \Phi_X  ~,
\end{align}
\(\tilde{T} \equiv \tilde{g}_{\mu\nu}\tilde{T}^{\mu\nu}\) is the trace of the matter stress-energy tensor in the Jordan frame,
and \(\tilde{g}_{\mu\nu} \equiv \Phi^{-1} g_{\mu\nu}\) is the metric in the Jordan frame. 

The disformal relation \eqref{eq: Tcurly definition} arises due to the factor
\begin{align}
     \dfrac{\delta \tilde{g}_{\mu \nu}}{\delta g_{\rho \sigma}} = \dfrac{1}{\Phi} \delta^{\rho}_{\mu}\delta^{\sigma}_{\nu} + \dfrac{\Phi_X}{\Phi^2}  g_{\mu \nu} \varphi^\rho \varphi^\sigma  ~.
\end{align}

These equations can be written as Eqs.~\eqref{eq: scalar equation}-\eqref{eq: A B C definitions} by using the (disformal) relation \eqref{eq: Tcurly definition},
and by substituting Eq.~\eqref{eq: scalar equation 2} into Eq.~\eqref{eq: matter equation 2}.

%%%%%%%%%%%%%%%%%%%%%%%%%%%%%%%%%%%%%%%%%%%%%%%%%%%
%%%%%%%%%%%%%%%%%%%%%%%%%%%%%%%%%%%%%%%%%%%%%%%%%%%
%%%%%%%%%%%%%%%%%%%%%%%%%%%%%%%%%%%%%%%%%%%%%%%%%%%

%

\bibliographystyle{utphys}
\bibliography{bibliography}% Produces the bibliography via BibTeX.

\end{document}